\documentclass[prl,showpacs,amsmath,amssymb,superscriptaddress,twocolumn]{revtex4}

\usepackage{amssymb}
\usepackage{amsmath}
\usepackage{amsfonts}
\usepackage{bm}
\usepackage{graphicx}
\usepackage{color}
\usepackage[usenames,dvipsnames]{xcolor}
\usepackage{dcolumn}
\usepackage{bm}
\usepackage{multirow}
\usepackage{array}

\begin{document}

\newcommand{\spp}{SrPt$_3$P}
\newcommand{\tc}{$T_{c}$}

\title{Lattice dynamical properties of superconducting SrPt$_3$P studied via inelastic x-ray scattering and density functional perturbation theory}

\author{D. A. Zocco}
\email[Corresponding author: ]{diego.zocco@kit.edu}
\affiliation{Institute for Solid State Physics (IFP), Karlsruhe Institute of Technology, D-76021 Karlsruhe, Germany}
\author{S. Krannich}
\affiliation{Institute for Solid State Physics (IFP), Karlsruhe Institute of Technology, D-76021 Karlsruhe, Germany}
\author{R. Heid}
\affiliation{Institute for Solid State Physics (IFP), Karlsruhe Institute of Technology, D-76021 Karlsruhe, Germany}
\author{K.-P. Bohnen}
\affiliation{Institute for Solid State Physics (IFP), Karlsruhe Institute of Technology, D-76021 Karlsruhe, Germany}
\author{T. Wolf}
\affiliation{Institute for Solid State Physics (IFP), Karlsruhe Institute of Technology, D-76021 Karlsruhe, Germany}
\author{T. Forrest}
\affiliation{European Synchrotron Radiation Facility (ESRF), F-38043 Grenoble Cedex, France}
\author{A. Bosak}
\affiliation{European Synchrotron Radiation Facility (ESRF), F-38043 Grenoble Cedex, France}
\author{F. Weber}
\email[Corresponding author: ]{frank.weber@kit.edu}
\affiliation{Institute for Solid State Physics (IFP), Karlsruhe Institute of Technology, D-76021 Karlsruhe, Germany}

\begin{abstract}
We present a study of the lattice dynamical properties of superconducting \spp\ (\tc\,=\,8.4\,K) via high-resolution inelastic x-ray scattering (IXS) and \textit{ab initio} calculations. Density functional perturbation theory including spin-orbit coupling results in enhanced electron-phonon coupling (EPC) for the optic phonon modes originating from the Pt(I) atoms, with energies $\sim$\,5\,meV, resulting in a large EPC constant $\lambda$\,$\sim$\,2. An overall softening of the IXS powder spectra occurs from room to low temperatures, consistent with the predicted strong EPC and with recent specific-heat experiments (2$\Delta_0$/$k_{\mathrm{B}}T_c$\,$\sim$\,5). The low-lying phonon modes observed in the experiments are approximately 1.5\,meV harder than the corresponding calculated phonon branch. Moreover, we do not find any changes in the spectra upon entering the superconducting phase. We conclude that current theoretical calculations underestimate the energy of the lowest band of phonon modes indicating that the coupling of these modes to the electronic subsystem is overestimated.
\end{abstract}


\maketitle

The search for high-temperature superconductivity has been lately focused in materials presenting unconventional Cooper-pairing mechanisms, such as in the copper-oxide and iron-based superconductors, in which it is believed that the attractive binding force between electrons arises from magnetic exchange interactions \cite{basov11}. Phonon-mediated superconductivity, however, has never left the scene, with materials like MgB$_2$ \cite{nagamatsu01} or hydrogen metallic alloys, in which it has been predicted \cite{ashcroft04} and recently demonstrated in high-pressure experiments\cite{eremets15} that high phonon frequencies due to the light atoms having strong electron-phonon coupling (EPC) can lead to high values of $T_c$. Additionally, low-frequency phonons in compounds containing heavier elements could give sizeable \tc\ values via enhanced EPC.

\begin{figure}[th]
\includegraphics[width=3.4in]{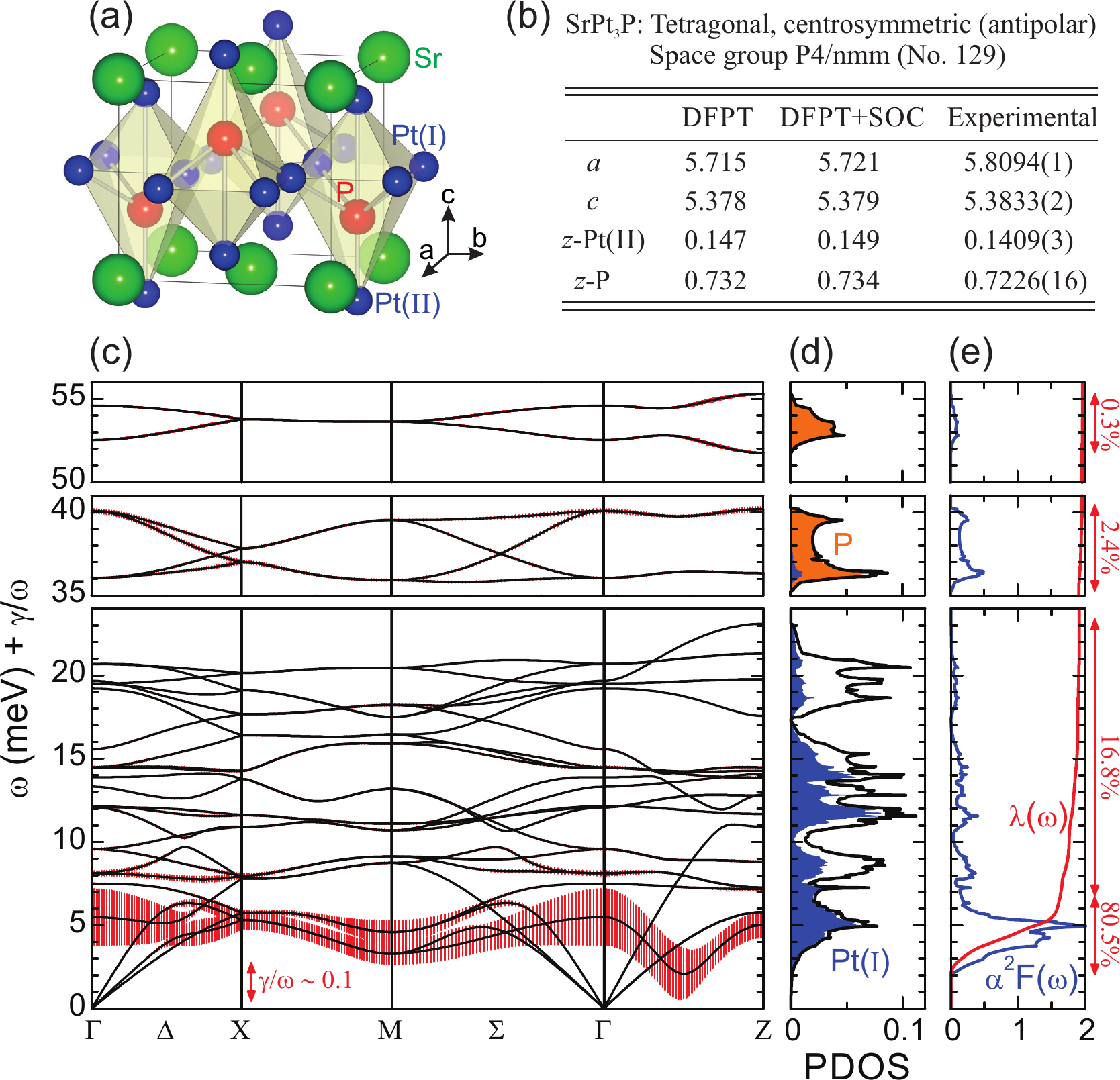}
\caption{(color online). (a) Crystal structure of \spp. Platinum atoms occupy two distinct positions, Pt(I) and (II). (b) Calculated and experimental \cite{takayama12} lattice parameters. (c) Calculated phonon dispersions $\omega(k)$ and relative phonon linewidths $\gamma / \omega$ (red vertical bars, amplified for clarity). SOC was included in the calculations (see text). (d) Calculated PDOS (black lines) vs $\omega$. The colored regions emphasize the dominant atoms determining the dispersion at low (Pt(I), blue) and high (P, orange) energies. (e) Calculated isotropic Eliashberg function $\alpha^2F(\omega)$ (blue line) and EPC constant $\lambda(\omega)$ (red line) vs $\omega$. Percentages indicate the contribution of $\alpha^2F(\omega)$ to $\lambda(\omega)$ in different energy ranges.}
\label{fig1}
\end{figure}

Recently, a new family of ternary platinum phosphide superconductors $A$Pt$_3$P with $A$\,=\,La, Ca, Sr was discovered \cite{takayama12}, presenting $T_c$'s of 1.5, 6.6 and 8.4\,K, respectively. These materials crystallize in a tetragonal antiperovskite structure (space group $P$4/$nmm$) similar to that of the non-centrosymmetric heavy-fermion superconductor CePt$_3$Si, although the phosphorous compounds maintain the inversion symmetry. The structure consists of alternating layers of $A$ atoms and distorted Pt$_6$P octahedra, which leads to two different Pt lattice sites (Fig.\,\ref{fig1}(a)). For \spp, specific-heat \cite{takayama12} $C$($T$) and muon-spin rotation \cite{khasanov14} ($\mu$SR) measurements revealed superconducting gap energies $\Delta_0$\,=\,1.85 and 1.55\,meV and ratios 2$\Delta_0$/$k_{\mathrm{B}}T_c$\,$\sim$\,5.1 and 4.3, respectively, providing evidence for very strong-coupling, fully-gapped $s$-wave superconductivity. Anomalous curvature of the upper critical field was interpreted as strong spin-orbit coupling (SOC) inherent to the 5$d$ Pt orbitals \cite{takayama12} but also as arising from a two-band superconducting state with equal gaps but differing values of the coherence lengths \cite{khasanov14}. 

The  coupling of superconducting charge carriers to low-lying phonons in \spp\ was first inferred from a non-linear normal-state specific heat \cite{takayama12} and it has been later corroborated by first-principle calculations \cite{nekrasov12,chen12,kang13,subedi13}. These initial theoretical studies reported a negligible effect of the SOC of the 5$d$-Pt states on superconductivity \cite{chen12,kang13,subedi13}. We present the first experimental study of the phonon spectrum of \spp\ performed via high-energy-resolution inelastic x-ray scattering (IXS) measurements of a powder sample, combined with theoretical calculations of the dynamical properties including SOC. An overall softening of the spectrum at low temperatures is consistent with the strong EPC obtained from our \textit{ab initio} calculations. The experimental results suggest, however, that theoretical calculations overestimate the strength of the EPC. Moreover, we show that the SOC effect strongly renormalizes the EPC constant $\lambda$ and that it must be taken into account for a comprehensive analysis of the data.

Density-functional theory (DFT) calculations were performed within a mixed-basis pseudopotential (PP) framework \cite{louie79,meyer97}, using the local density approximation (LDA) in the parametrization of Perdew-Wang \cite{perdew92}. Norm-conserving PPs were constructed from all-electron relativistic atom calculations according to the scheme of Vanderbilt \cite{vande85} and include non-linear core corrections. SOC was incorporated within the PP approach \cite{kleinman80,bachelet82} and treated fully self-consistently in the ground-state calculations \cite{heid10}. Plane waves up to a kinetic energy of 22\,Ry were augmented with local functions of $s$ and $p$ type at the P sites, and $s$, $p$, and $d$ types at Sr and Pt sites. This choice of the basis set guaranteed sufficient convergence of electronic and phononic properties. Brillouin zone summations were performed with regular $k$-point meshes in combination with the standard smearing technique \cite{fu83} employing a Gaussian broadening of 0.1\,eV \cite{note_mesh}. The results for the lattice optimizations are summarized in Fig.\,\ref{fig1}(b). They indicate a small over-binding of the in-plane lattice constant $a$ by 1.5\% typical for LDA, while $c$ matches well the experimental value. For comparison, previous structure optimization based on the generalized gradient approximation resulted in lattice constants exceeding the experimental values by about 1.5\% for both $a$ and $c$ directions \cite{subedi13}. Overall, our LDA band structure calculations including SOC are in good agreement with the previously reported results, with bands near the Fermi level dominated by Pt(I) and Pt(II) 5$d$-states \cite{chen12,kang13,subedi13}.

Phonon and electron-phonon coupling (EPC) properties were obtained via density-functional perturbation theory (DFPT) \cite{baron01,heid99}. The calculated phonon dispersion $\omega(k)$ of \spp\ and the phonon density of states (PDOS) are displayed in Figures\,\ref{fig1}(c) and (d), respectively. At high frequencies, the phonon dispersion is dominated by the vibrations of P atoms, which determines almost entirely the PDOS. On the other hand, optic modes originating from the in-plane motion of Pt(I) atoms dominate the low-frequency part of the phonon dispersion and PDOS. The calculations show that the EPC is strongly manifested in the phonon modes with energies $\sim$\,5\,meV. This strong coupling is represented \cite{allen72} by the relative phonon linewidth $\gamma / \omega$ (red vertical bars in Fig.\,\ref{fig1}(c))
\small
\begin{equation}\label{linewidth}
\frac{\gamma_{\mathbf{q},\lambda}}{\omega_{\mathbf{q},\lambda}} = 2 \pi \sum_{\mathbf{k},\nu,\nu'}\vert \text{g}_{\mathbf{k}+\mathbf{q},\nu';\mathbf{k},\nu}^{\mathbf{q},\lambda}\vert^2 \delta(\epsilon_{\mathbf{k},\nu}{-}E_{\mathrm{F}})\delta(\epsilon_{\mathbf{k}+\mathbf{q},\nu'}{-}E_{\mathrm{F}})
\end{equation}
\normalsize
where $\epsilon_{\mathbf{k},\nu}$ are the electronic energies with momentum $\mathbf{k}$ and band index $\nu$, $\text{g}_{\mathbf{k}+\mathbf{q},\nu';\mathbf{k},\nu}^{\mathbf{q},\lambda}$ are the electron-phonon matrix elements, and $E_{\mathrm{F}}$ is the Fermi energy. The relative linewidth is a measure of the coupling strength of an individual phonon mode, and enters directly in the Eliashberg function for an isotropic $s$-wave superconductor
\begin{equation}\label{eliashberg}
\alpha^2F(\omega) = \frac{1}{2\pi\hbar N(E_{\mathrm{F}})}\sum_{\mathbf{q},\lambda} \frac{\gamma_{\mathbf{q},\lambda}}{\omega_{\mathbf{q},\lambda}}\delta(\omega-\omega_{\mathbf{q},\lambda})\text{,}
\end{equation}
where $N(E_{\mathrm{F}})$ denotes the electronic density of states per atom and spin at the Fermi energy, and from which one can estimate the average EPC constant $\lambda$:
\begin{equation}\label{lambda}
\lambda = 2 \int_0^{\infty} d\omega \frac{\alpha^2F(\omega) }{\omega} = \frac{1}{\pi\hbar N(E_{\mathrm{F}})}\sum_{\mathbf{q},\lambda} \frac{\gamma_{\mathbf{q},\lambda}}{\omega_{\mathbf{q},\lambda}^2}.
\end{equation}
\noindent The two functions are plotted in Fig.\,\ref{fig1}(e). Clearly, the strongly-coupled low-energy Pt(I) vibrations constitute the main contribution to $\alpha^2F(\omega)$, resulting in the large coupling constant value $\lambda$\,=\,1.97, and transition temperature \tc\,=\,9.4\,K estimated by solving the Eliashberg-gap equations using a Coulomb pseudopotential $\mu^{*}$\,=\,0.1. For \spp\ the strong coupling of the low-lying phonons to the electronic states is consistent with the experimental observations made by Takayama \textit{et al.} from electrical resistivity and specific-heat measurements \cite{takayama12}. The inclusion of SOC in our calculations is responsible for a 23\% increase in the value of $\lambda$ (from 1.60 to 1.97), mainly due to a shift of the low-energy phonons towards lower energies (Eq.\,[\ref{lambda}]). SOC was recently included in the calculations of $\lambda$ for elementary Pb, resulting in larger values that explain the strong EPC effects observed in experiments \cite{heid10}. We conclude that SOC effects must be taken into account for a comprehensive theoretical treatment of \spp.
\begin{figure}[t]
\includegraphics[width=3.4in]{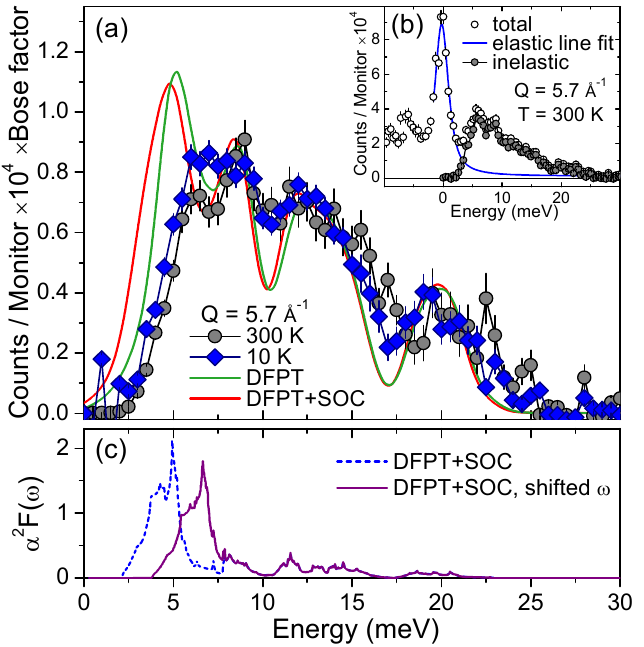}
\caption{(color online). (a) Inelastic data from combined energy scans at $Q$\,=\,5.7\,\AA$^{-1}$ obtained at 300\,K (circles) and 10\,K (diamonds). Powder-average calculations of the dynamical structure factors are shown for $Q$\,=\,5.7\,\AA$^{-1}$ as green (no SOC) and red (with SOC) lines. (b) A representative energy scan measured between $-$10\,meV and 30\,meV (empty symbols). The inelastic phonon component (full symbols) was obtained after subtracting the fit to the elastic peak contribution (blue line). (c) Scaling of the Eliashberg function $\alpha^2F(\omega)$ by a shift in energy $\omega_{sh}$\,$\rightarrow$\,$\omega+1.7$\,meV.}
\label{fig2}
\end{figure}

High-energy-resolution inelastic x-ray scattering experiments were performed on polycrystalline powder samples  of \spp\ prepared via conventional solid state reaction. X-ray diffraction performed on our samples revealed powders containing more than 90\% of the desired \spp\ superconducting phase, with a  \tc\,=\,8.5\,K determined from ac-susceptibility measurements. The IXS measurements were performed at the ID28 beamline located at the European Synchrotron Radiation Facility (ESRF), between room temperature and 2.5\,K using a continuous flow cryostat. A monochromatic incident beam with energy $E_i$\,=\,21.747\,keV was obtained from the (11,11,11)-Bragg reflection of a silicon-crystal monochromator which operates in a backscattering geometry. The energy resolution in this experiment was $\sim$\,1.5\,meV \cite{note_ixs}.

The energy scans performed for $Q$\,=\,5.7\,\AA$^{-1}$ at 300\,K and 10\,K are displayed in Fig.\,\ref{fig2}(a). Fig.\,\ref{fig2}(b) displays an example of raw data (empty circles) measured at 300\,K. The spectrum consist mainly of a strong elastic peak and the inelastic contribution originating from phonon scattering. Tails of Debye-Scherrer rings were largely avoided by a suitable choice of the scattering angle. The elastic peaks were fitted with a pseudo-Voight function (blue line) with 80\,\% of Lorentzian-shape component, consistent with the spectrometer's resolution characterization. Inelastic phonon spectra (filled circles) were obtained after subtracting the elastic peak from the total counts and were normalized with the corresponding Bose-factor. As Fig.\,\ref{fig2}(b) shows, the low energy part of the spectra is dominated by the temperature independent elastic scattering. Therefore the inelastic component could not be analyzed reliably for $E$\,$\leq$\,2\,meV.
\begin{figure}[t]
\includegraphics[width=3.4in]{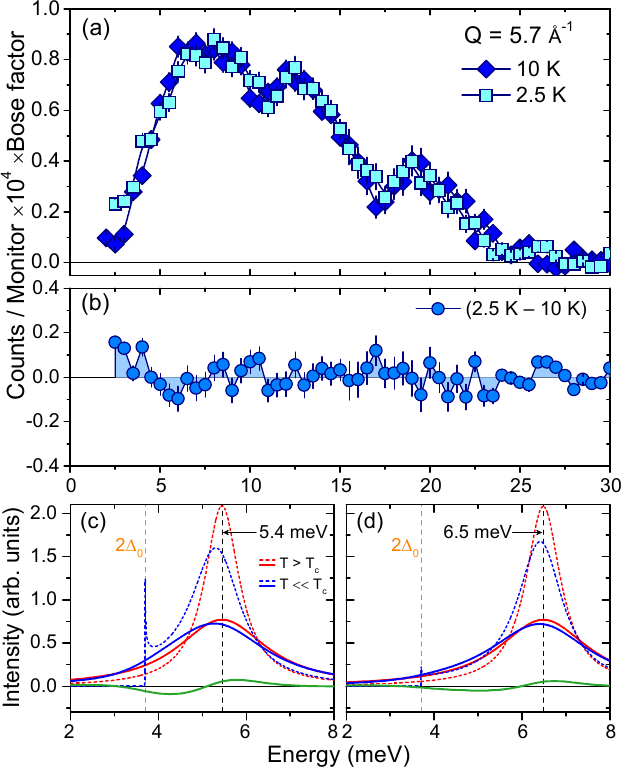}
\caption{(color online). (a) Inelastic data from combined energy scans at $Q$\,=\,5.7\,\AA$^{-1}$ obtained above (diamonds) and below (squares) \tc. The difference between these spectra is displayed in (b). (c-d) Example of calculations of phonon line shapes based on the theory of Allen \textit{et\,al.} \cite{allen97} for the first optic phonon mode at the $\Gamma$ point for a single crystal. Red and blue curves correspond to temperatures above and below \tc, respectively. The broadening of the dashed curves is due only to the EPC effect presented in Fig.\,\ref{fig1}(c), while for the solid lines, an additional instrumental broadening is also included (see text).} 
\label{fig3}
\end{figure}

An overall softening of the phonon spectrum is observed when cooling from 300\,K to 10\,K. This is contrary to what is expected, that is, a stiffer lattice yielding larger phonon frequencies. However, an enhancement of the electronic density of states at the Fermi level upon cooling can cause damping of lattice vibrations in systems where EPC is important, as has been observed, for example, in Nb and YNi$_2$B$_2$C \cite{waka86,weber14}. In particular, the 10\,K data shows a pronounced increase of spectral weight below 10\,meV with respect to the room-temperature data, in agreement with the enhanced line broadening and the strong coupling of these low-energy phonons to the electronic system predicted by the DFPT calculations.

To compare the experimental results with the predictions of the DFPT, calculations of inelastic powder spectra including experimental and EPC line-broadening (from DFT calculations) are presented for $Q$\,=\,5.7\,\AA$^{-1}$ in Fig.\,\ref{fig2}(a) as green (no SOC) and red (with SOC) lines \cite{note_dynstrucfact}. The calculations match fairly well the energy and relative intensity of the peaks of the experimental data above 7\,meV. However, there is a strong discrepancy between experiment and theory at lower energies. First, the position of the calculated low-energy phonon band lies about 1.7\,meV below the lowest measured peak. Second, the amplitude of the calculated phonon powder spectra differs from the experimental data particularly at low energies. The inclusion of SOC enhances the EPC and further softens the powder spectrum (red line), leading to a slightly smaller amplitude. Still, the inclusion of SOC in the calculations does not provide a good agreement with the experimental measurement.

We conclude that the DFPT calculations underestimate the energy of the lowest band of phonon modes indicating that the coupling of these modes to the electronic subsystem is overestimated. If the calculations were adjusted to increase the energy of this phonon band, then this would result automatically in a decrease in the strength of the estimated EPC. In order to simulate the impact of such a shift on the EPC and superconducting properties, we artificially modified the Eliashberg function by shifting $\omega_{sh}$ \,=\, $\omega + 1.7$\,meV in an energy range up to 6.2\,meV. The resulting scaled Eliashberg function $\alpha^2F(\omega_{sh})$ \,=\, $\left(\omega / \omega _{sh} \right) \alpha^2F(\omega)$ is presented in Fig.\,\ref{fig2}(c). This procedure yields a reduction of the EPC parameter $\lambda$ from 1.97 to 1.24, corresponding to an estimated \tc\,=\,8\,K for $\mu^{*}$\,=\,0.1, in good agreement with the experiment. This analysis highlights the uncertainties in predicting an accurate value for the coupling constant in \spp. The fact that the most complete \textit{ab initio} calculation including SOC and an adequate selection of $k$ and $q$-point meshes \cite{note_mesh} resulted in the largest underestimation of the frequency of the low-frequency phonon mode, emphasizes the accuracy limits of current DFT approximations when dealing with soft phonons. A high sensitivity on calculational details is quite often found for materials where soft modes exist, as occurs for example in charge-density-wave systems \cite{weber11, bianco15}. If such modes do contribute significantly to the electron-phonon coupling, the effect of the aforementioned sensitivity is further amplified in the prediction for $\lambda$. We note, however, that in contrast to $\lambda$, \tc\ is much less affected by this uncertainty, since it depends on both the strength of the EPC and frequency of the phonon mode dominating the superconducting pairing.

Upon entering the superconducting state, a change in the phonon spectra at energies near 2$\Delta$($T$) is expected to occur in systems presenting strong EPC \cite{allen97, weber08}, because the number of possible excited electronic states by a phonon is zero for energies smaller than 2$\Delta$. Fig.\,\ref{fig3}(a) displays the combined spectra collected above (10\,K) and below (2.5\,K) \tc\ and, in (b), their difference. The inelastic data show no clear change between the two measurements at low energies near 2$\Delta_0$\,$\sim$\,3.7\,meV within statistical error. The calculations of phonon line shapes based on the theory of Allen \textit{et\,al.} \cite{allen97} are presented in Fig.\,\ref{fig3}(c) and (d) for the single-crystalline case, and suggest two effects that are detrimental to the experimental observation of a change in the phonon spectrum below \tc\ in our measurements. First, in Fig.\,\ref{fig3}(c), a phonon peak at 5.4\,meV (dashed red line), as predicted by DFPT at the $\Gamma$ point with a FWHM of 1\,meV due to EPC, would soften and develop a sharp anomaly near $E$\,=2$\Delta_0$ for $T$\,$\ll$\,\tc\ (dashed blue line). When one considers the experimental resolution (solid red and blue lines, pseudo-Voight line shape with a FWHM of 1.55\,meV and 80\% Lorentzian weight), only a mild softening would be noticeable at low temperatures, resulting in a small difference between the curves above and below \tc\ (green line). Second, a phonon peak shifted to higher energies as it is found in the IXS data would make the effect even smaller, as shown in Fig.\,\ref{fig3}(d). The effect below \tc\ is expected to be less prominent in the powder experiments due to $Q$-space averaging and to the larger background arising from elastic scattering at these lower energy transfers.

In summary, we presented IXS experiments performed on powder samples of \spp. The room-temperature phonon spectrum shifts towards lower energies upon cooling, which can be understood in terms of a damping of lattice vibrations due to strong EPC and the enhanced electronic density of states near the Fermi level at low temperatures. A strong EPC is predicted by our calculations, in close agreement with previously reported results. The inclusion of SOC due to the presence of Pt-5$d$ states yields a large EPC coupling constant $\lambda$\,$\sim$\,2. However, when comparing the experimental and theoretical data with and without SOC, we conclude that theoretical models overestimate the coupling of the low-energy phonon modes to the electronic system. Adjusted EPC parameters reproduce the experimentally observed \tc\ fairly well. Further experiments should be performed, if possible, in currently unavailable single crystals, to explore the low-energy phonon spectrum with regard to the intrinsic phonon linewidths as well as the superconductivity-induced changes of the line shape.

D.Z., S.K. and F.W. acknowledge supported by the Helmholtz Association under contract VH-NG.840.


\begin{thebibliography}{30}
\bibitem[{not({\natexlab{a}})}]{note_mesh}
\bibinfo{note}{Due to the large couplings, phonon frequencies and EPC exhibited
  a strong dependence on the applied broadening and the density of the $k$
  mesh. Sufficient convergence was obtained with a 12$\times$12$\times$12 $k$
  mesh for the phonon dispersion, and a 16$\times$16$\times$16 $k$ mesh for the
  calculation of the EPC properties. Phonon and EPC properties were determined
  via DFPT on a 4$\times$4$\times$4 $q$-point mesh including contributions from
  SOC \cite{heid10}. These results were then interpolated for arbitrary $q$
  points using Fourier techniques to accurately determine the Eliashberg
  function. Superconducting properties were analyzed by solving the isotropic
  Eliashberg-gap equations.}
\bibitem[{not({\natexlab{b}})}]{note_ixs}
\bibinfo{note}{Typical counting times varied from 60 to 120 seconds using
  0.5\,meV energy steps. Diffraction scans (2$\theta$) were performed before
  measurements to determine the position of elastic peaks. The energy scans
  displayed in Figs.\,\ref{fig2} and \ref{fig3} correspond to the selected
  angle of $\sim$\,30$^{\circ}$ ($Q$\,=\,5.7\,\AA$^{-1}$) with low number of
  elastic counts, which matched the position of the instrument's analyzer
  No.\,2.}
\bibitem[{not({\natexlab{c}})}]{note_dynstrucfact}
\bibinfo{note}{Phonon structure factors (PSF) were calculated for a single
  crystal of SrPt$_3$P, including the experimental resolution, for a total of
  853 Brillouin zones with scattering vectors of absolute value
  $Q$\,$\leq$\,12\,\AA$^{-1}$ in the first octant of the reciprocal space and
  for the respective 8000 reduced wavevectors. To account for the
  1/$\omega$-dependence of the phonon intensity, the PSF have been weighted in
  accordance with the relevant phonon energy. For the powder averaging, the
  individual PSF were summed within spherical shells $Q+dQ$. Due to the
  quadratic increasing number of $Q$-points, the $Q^2$-dependence for the
  powder spectrum in this summation is implied.}
\end{thebibliography}
\end{document}